\documentclass[conference]{IEEEtran}
\IEEEoverridecommandlockouts

\usepackage{cite}
\usepackage{amsmath,amssymb,amsfonts}
\usepackage{algorithmic}
\usepackage{graphicx}
\usepackage{textcomp}
\usepackage{xcolor}
\usepackage{multirow}
\usepackage{float}
\usepackage{graphicx}
\usepackage{subcaption}
\usepackage{algorithm}
\usepackage{float}
\usepackage{booktabs}
\usepackage{url}
\usepackage{enumitem}
\usepackage{placeins}
\usepackage{array} 

\def\BibTeX{{\rm B\kern-.05em{\sc i\kern-.025em b}\kern-.08em
    T\kern-.1667em\lower.7ex\hbox{E}\kern-.125emX}}
\begin{document}

\newcommand\aat[1]{{\color{blue} \bf [Tanvir: #1]}}

\newcommand\xz[1]{{\color{orange} \bf [Zhong: #1]}}

\title{Combining Textual and Spectral Features for Robust Classification of Pilot Communications\\
\thanks{This work was supported by the Nebraska Research Initiative through the University of Nebraska Collaboration Initiative.}
}

\author{
    \IEEEauthorblockN{
    Abdullah All Tanvir\IEEEauthorrefmark{1},
    Chenyu Huang\IEEEauthorrefmark{2},
    Moe Alahmad\IEEEauthorrefmark{3},
    Chuyang Yang\IEEEauthorrefmark{4},
    Xin Zhong\IEEEauthorrefmark{1}}\
    
    \smallskip
    \IEEEauthorblockA{
    \IEEEauthorrefmark{1}
    Department of Computer Science, University of Nebraska Omaha, Omaha, NE, USA
    \\ 
    \IEEEauthorrefmark{2}
    Aviation Institute, University of Nebraska Omaha, Omaha, NE, USA
    \\
    \IEEEauthorrefmark{3}
    Durham School of Architectural Engineering \& Construction, 
    University of Nebraska Lincoln, Lincoln, NE, USA
    \\
    \IEEEauthorrefmark{4}
    School of Graduate Studies, Embry-Riddle Aeronautical University Daytona Beach, Daytona Beach, FL, USA
    \\
    atanvir@unomaha.edu, chenyuhuang@unomaha.edu, malahmad2@unl.edu, yangc11@erau.edu, xzhong@unomaha.edu
    }
}


\maketitle

\begin{abstract}
Accurate estimation of aircraft operations, such as takeoffs and landings, is critical for effective airport management—yet remains challenging, especially at non-towered facilities lacking dedicated surveillance infrastructure. This paper presents a novel dual-pipeline machine learning framework that classifies pilot radio communications using both textual and spectral features. Audio data collected from a non-towered U.S. airport was annotated by certified pilots with operational intent labels and preprocessed through automatic speech recognition and Mel-spectrogram extraction. We evaluate a wide range of traditional classifiers and deep learning models—including ensemble methods, LSTM, and CNN across both pipelines. To our knowledge, this is the first system to classify operational aircraft intent using a dual-pipeline ML framework on real-world air traffic audio. Our results demonstrate that spectral features combined with deep architectures consistently yield superior classification performance, with F1-scores exceeding 91\%. Data augmentation further improves robustness to real-world audio variability. The proposed approach is scalable, cost-effective, and deployable without additional infrastructure, offering a practical solution for air traffic monitoring at general aviation airports.
\end{abstract}

\vspace{-1.0em}

\section{Introduction}
\vspace{-0.5em}




Accurate monitoring of aircraft operations is essential to the strategic functioning of airports, yet remains challenging—especially for non-towered facilities. Daily and annual counts of takeoffs and landings are critical for both towered and non-towered airports, supporting a wide range of airport management tasks such as strategic planning, environmental assessments, capital improvement programs, funding justification, and personnel allocation.
Insights derived from operational counts can substantially inform decisions related to airport expansions, infrastructure upgrades, and policy formulation. In the United States, only 521 of the 5,165 public-use airports are staffed with air traffic control personnel capable of tracking aircraft movements, underscoring a considerable gap in operational data coverage~\cite{faa2024airtraffic}.

At towered airports, operational aircraft counts are typically recorded by air traffic control (ATC) towers, though these data often lack details and completeness. Many control towers operate on a part-time basis, leading to missed aircraft activity and resulting in incomplete operational records. 
In response to these limitations, the Federal Aviation Administration (FAA), in collaboration with the aviation industry, has undertaken various initiatives in recent years to improve the estimation of aircraft operations. A wide array of methods has been employed at both towered and non-towered airports, leveraging technologies such as acoustic sensors, airport visitor logs, fuel sales data, video image detection systems, aircraft transponders, and statistical modeling techniques \cite{nase2007counting, NAP22182, mott2019evaluation, yang2019technology}. Despite these efforts, existing technologies remain constrained by high costs, limited adaptability, and inconsistent accuracy, failing to offer a universally applicable and economical solution for all airport types. 
This challenge is particularly acute for the nation’s general aviation airports, which collectively service over 214,000 aircraft and account for more than 28 million flight hours annually across more than 5,100 U.S. public airports in the United States \cite{gama2025contribution}. 
The absence of a reliable, scalable, and cost-effective approach to accurately monitoring aircraft operations underscores the urgent need for innovative solutions. Addressing this data gap is essential to enhancing decision-making in airport planning, infrastructure development, and policy formulation. 

From a machine learning standpoint, pilot communication audio offers a valuable yet underutilized data source. Unlike physical sensors, these recordings are already widespread at airports and contain rich operational information. However, challenges such as unstructured language, overlapping speech, background noise, and limited labeled data make modeling and large-scale supervised learning difficult.

To address this, we propose a classification framework that leverages both textual and spectral features from air traffic communication. The textual pipeline applies automatic speech recognition (ASR) followed by TF-IDF vectorization, while the spectral pipeline extracts Mel-spectrograms to capture acoustic patterns. These features are used to train a range of models, including traditional classifiers, LSTMs, and CNNs.

Our contributions include: (1) a dual-modality machine learning framework that overcomes speech irregularity and background noise challenges by leveraging both textual and spectral representations of real-world pilot radio communications; (2) a structured data collection and augmentation framework that addresses the scarcity of labeled air traffic audio and enables robust model training under realistic conditions; and (3) the first application of this approach to infer operational intent (landing vs. takeoff) from air traffic communication at non-towered airports. 
As pilot communication audio is already widely available, our framework requires no additional hardware and is deployable at scale, making it a cost-effective and practical solution for aviation monitoring.


\section{Related Work}
\label{sec:related_work}
\vspace{-0.25em}
This section reviews prior work on aviation audio collection and machine learning-based audio classification, covering data acquisition, preprocessing, feature extraction, and classification methods. It also highlights recent deep learning advances for tasks like cockpit audio interpretation and anomaly detection.
\vspace{-0.25em}
\subsection{\textbf{Aircraft Operation Estimation Approaches}}
A variety of methods have been developed to estimate aircraft operations, particularly at airports lacking full-time ATC towers. Among these, aircraft transponder signal analysis techniques have been extensively explored \cite{mott2017estimation, mott2017accuracy, farhadmanesh2022general}. Transponder-based methods leverage Mode S Extended Squitter (ES), Mode S, and Mode C signals, typically detected using software-defined radio (SDR) systems to infer aircraft proximity and movement. Adaptive Kalman filters are often employed to improve distance estimation accuracy. Transponder-based approaches offer the advantage of providing operational counts without the need for additional ground-based infrastructure. However, they require that aircraft be equipped with onboard transponders, limiting their applicability to cooperative traffic. Moreover, reported error rates vary by deployment condition, ranging from -10.2\% to +7.6\% \cite{yang2019technology}.

Other techniques, such as flight tracking and acoustic sensing, have also been used to estimate aircraft operations \cite{pretto2024aircraft}. Patrikar at el. introduced the TartanAviation multi-modal dataset to support airspace management in both towered and non-towered terminal areas \cite{patrikar2025image}. High-resolution flight tracking data and ground-based acoustic sensors support the reconstruction of low-altitude flight paths and operations. Nonetheless, these approaches face persistent challenges, including environmental noise interference, incomplete signal coverage, and data validation limitations. In particular, acoustic-based systems struggle to identify aircraft and often lack precision.

Manual and semi-automated methods remain in use, especially at non-towered airports. These include the use of indirect indicators such as fuel sales, visitor logs, and other administrative records. While straightforward, these methods are labor-intensive and offer limited accuracy.

A review of related work reveals a progression from basic manual counts to sophisticated technological solutions such as transponder-based monitoring and machine learning. Despite this evolution, a universally applicable, cost-effective, and scalable solution remains elusive. Each method has inherent trade-offs, and the diversity of airport environments necessitates adaptable approaches. There is a pressing need for innovative systems that can provide accurate, low-cost estimates suitable for both towered and non-towered airports, enabling better resource allocation, planning, and policy development.
\vspace{-0.25em}
\subsection{\textbf{Machine Learning for Aviation Audio Analysis}}
Audio classification plays a critical role in aviation environments, particularly for monitoring cockpit communication, detecting anomalies in air traffic control (ATC) transmissions, and enhancing situational awareness in airport operations. Unlike general audio tasks, aviation audio often contains overlapping speech, high ambient noise, and domain-specific terminology, making classification particularly challenging. As a result, robust preprocessing, noise reduction, and domain-adaptive modeling are essential. 

Recent research has explored the use of advanced machine learning and natural language processing techniques to tackle a range of operational and safety challenges within the aviation domain \cite{yang2023natural, alreshidi2024advancing, ohneiser2024text}. For instance, Chen et al. introduce the Audio Scanning Network (ASNet), a framework that harnesses rich audio features to enable stable and accurate audio classification \cite{chen2024audio}. Additionally, Castro-Ospina et al. investigate a graph-based approach to audio classification, demonstrating its potential in structured audio data analysis \cite{castro2024graph}. Among these efforts, automatic speech recognition (ASR) has emerged as one of the most actively explored areas, with a wide range of machine learning models applied to analyze audio communications between pilots and air traffic controllers for different downstream applications \cite{badrinath2022automatic, lin2019real, sun2021automatic, lin2019realair}. 

The development of machine learning models for air traffic communication introduces several domain-specific challenges. Most notably, the audio is highly unstructured, featuring domain-specific terminology, overlapping speech, variable accents, and significant background noise. These factors complicate both transcription and acoustic modeling. In addition, the scarcity of labeled datasets limits the application of large-scale supervised learning approaches. As a result, effective models must be robust to noisy and variable input while remaining data-efficient. Our work directly addresses these issues through a dual-pipeline classification framework that combines spectral and semantic representations and uses data augmentation to enhance generalization.


\section{Methodology}
\label{sec:methodology}

\begin{figure}[t!]
    \centering
    \vspace{-1.0em}
    \includegraphics[width=\columnwidth]{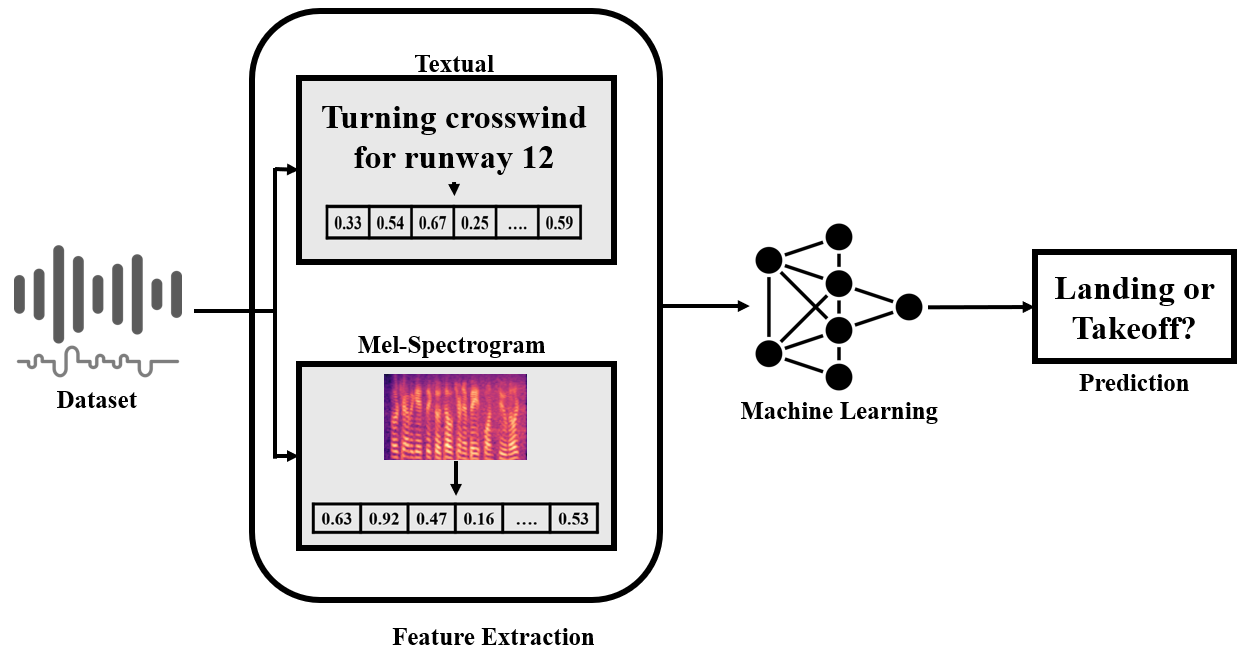}
    \vspace{-1.5em}
    \caption{Proposed dual-modality framework combining ASR-based text classification and Mel-spectrogram-based audio classification.}
    \label{fig:workflow}
    \vspace{-2.0em}
\end{figure}

To address the audio classification task, we adopt two complementary approaches: the Textual approach, which involves transcribing audio using ASR for text-based classification, and the Spectral approach, which extracts Mel-spectrograms directly from audio signals. Both traditional machine learning and deep learning models are applied to assess classification performance. The subsequent sections outline the dataset, preprocessing methods, model configurations, and evaluation criteria. Figure~\ref{fig:workflow} illustrates the overall idea of the proposed method. During inference, the two pipelines are used independently; predictions are generated separately for each, allowing for direct performance comparison without ensembling. This separation enables a clear understanding of the individual contribution of textual and spectral features.

\vspace{-0.5em}
\subsection{\textbf{Data Collection and Processing}}
\label{subsec:data}
To support machine learning classification of aircraft operational intent, we construct a domain-specific dataset from pilot radio communication recordings at a non-towered airport in Nebraska, United States. These recordings are captured over a three-month period using the Common Traffic Advisory Frequency (CTAF) and Universal Communications Frequency (UNICOM), resulting in over 68 hours of audio.

The raw audio is segmented into 2,489 distinct utterances based on clear pauses and transmission boundaries. These clips represent a wide range of real-world variations, including overlapping transmissions, variable audio quality, and diverse speaker accents. The data reflect aviation-specific language, typically expressed in compact and non-grammatical phrases optimized for rapid information exchange.

To enable supervised learning, each audio clip is manually annotated by three licensed pilots with extensive radio communication experience. The annotations includes:
(1) Operational Intent: ``Landing'' or ``Takeoff''; 
(2) Aircraft Position: e.g., ``downwind,'' ``base leg,'' ``final''; 
(3) Callsign: the aircraft’s tail number. 
Disagreements are resolved by majority voting, and uncertain clips were excluded. Table~\ref{tab:sample_data} shows sample examples for each label.

\begin{table}[h!]
\centering
\caption{Example data for landing and takeoff labels.}
\small
\begin{tabular}{|>{\centering\arraybackslash}p{7cm}|c|}
\hline
\textbf{Text} & \textbf{Label} \\
\hline
Turning crosswind for runway 12. & Landing \\
\hline
Departing runway 12 and staying in the pattern. & Takeoff \\
\hline
Reduce speed and descend to 3000 feet for landing. & Landing \\
\hline
Taxi into position and hold for takeoff. & Takeoff \\
\hline
\end{tabular}
\label{tab:sample_data}
\end{table}


This annotated dataset represents a rare, structured resource for training machine learning models to infer intent from unstructured aviation audio. It is designed to support both textual and spectral feature extraction pipelines, enabling dual-modality learning under challenging acoustic conditions.



\subsection{\textbf{Textual Feature Extraction with Spectral Subtraction}}
\label{subsec:audio_text}


Let \( x(t) \) denote the time-domain audio signal. To reduce background noise, we apply a spectral subtraction technique. The signal is first converted to the frequency domain using the Short-Time Fourier Transform (STFT):
\vspace{-0.5em}
\begin{equation}
X(t, f) = \text{STFT}\{x(t)\}
\end{equation}

The noise power spectral density (PSD) is estimated from the first \( T \) frames of the signal, \( x_n(t) \), assuming they contain only background noise:
\vspace{-0.5em}
\begin{equation}
|N(f)| = \frac{1}{T} \sum_{t=1}^{T} |X_n(t, f)|
\end{equation}

Denoising is then performed via spectral subtraction:
\vspace{-0.5em}
\begin{equation}
|\hat{X}(t, f)| = \max\left(|X(t, f)| - |N(f)|, 0\right)
\end{equation}

To preserve speech quality, a temporal smoothing filter is applied across frames. The enhanced signal is reconstructed using the inverse STFT (ISTFT) and the original phase \( \angle X(t, f) \):
\vspace{-0.5em}
\begin{equation}
\hat{x}(t) = \text{ISTFT}\left(|\hat{X}(t, f)| \cdot e^{j\angle X(t, f)}\right)
\end{equation}

The cleaned signal \( \hat{x}(t) \) is transcribed into text using the Google Web Speech API~\cite{googleASR}, resulting in a sequence of words \( w_1, w_2, \dots, w_n \). These transcriptions are transformed into structured features using Term Frequency–Inverse Document Frequency (TF-IDF) vectorization. The TF-IDF value for a term \( t \) in document \( d \) is computed as:
\vspace{-0.5em}
\begin{equation}
\text{TF-IDF}(t, d) = \text{tf}(t, d) \cdot \log\left(\frac{N}{\text{df}(t)}\right)
\end{equation}

where \( \text{tf}(t, d) \) is the term frequency of term \( t \) in document \( d \), \( \text{df}(t) \) is the number of documents containing \( t \), and \( N \) is the total number of documents.

This process yields a sparse feature vector \( \mathbf{v}_d \in \mathbb{R}^m \), where \( m \) is the size of the vocabulary. These numerical features are then used as input to machine learning models for classification. TF-IDF was selected for its ability to emphasize informative, aviation-specific terminology while down-weighting common words, making it ideal for sparse, domain-specific text data.

\subsection{\textbf{Spectral Feature Extraction with Mel-Spectrograms}}
\label{subsec:direct_audio}

For the direct audio-based classification pipeline, we extracted Mel-spectrogram representations from each audio recording to serve as input to the model. Each audio file was first resampled to a standard sampling rate of 22{,}050 Hz and truncated or zero-padded to a fixed duration of 3 seconds to ensure consistency.

Let \( x(t) \) denote the time-domain signal. We applied a 2048-point Fast Fourier Transform (FFT) with a hop length of 512 to compute the short-time magnitude spectrum. The signal was then mapped onto the Mel scale using a filter bank of \( M = 128 \) triangular filters, resulting in the Mel-spectrogram:
\vspace{-0.5em}
\begin{equation}
S(m, n) = \sum_{k=1}^{K} |X(k, n)|^2 \cdot H_m(k), \quad m = 1, 2, \dots, M
\end{equation}

where \( X(k, n) \) is the FFT of the \( n \)-th frame at frequency bin \( k \), and \( H_m(k) \) is the Mel filter bank. The resulting spectrogram \( S(m, n) \) was converted to the decibel (dB) scale:
\vspace{-0.5em}
\begin{equation}
S_{\text{dB}}(m, n) = 10 \cdot \log_{10}(S(m, n) + \epsilon)
\end{equation}

where \( \epsilon \) is a small constant added for numerical stability. The spectrograms were then min-max normalized to the \([0, 1]\) and reshaped to a fixed size of \( 128 \times 130 \) time-frequency bins.

\begin{figure}[!ht]
    \centering
    \vspace{-1.0em}
    \includegraphics[width=\columnwidth]{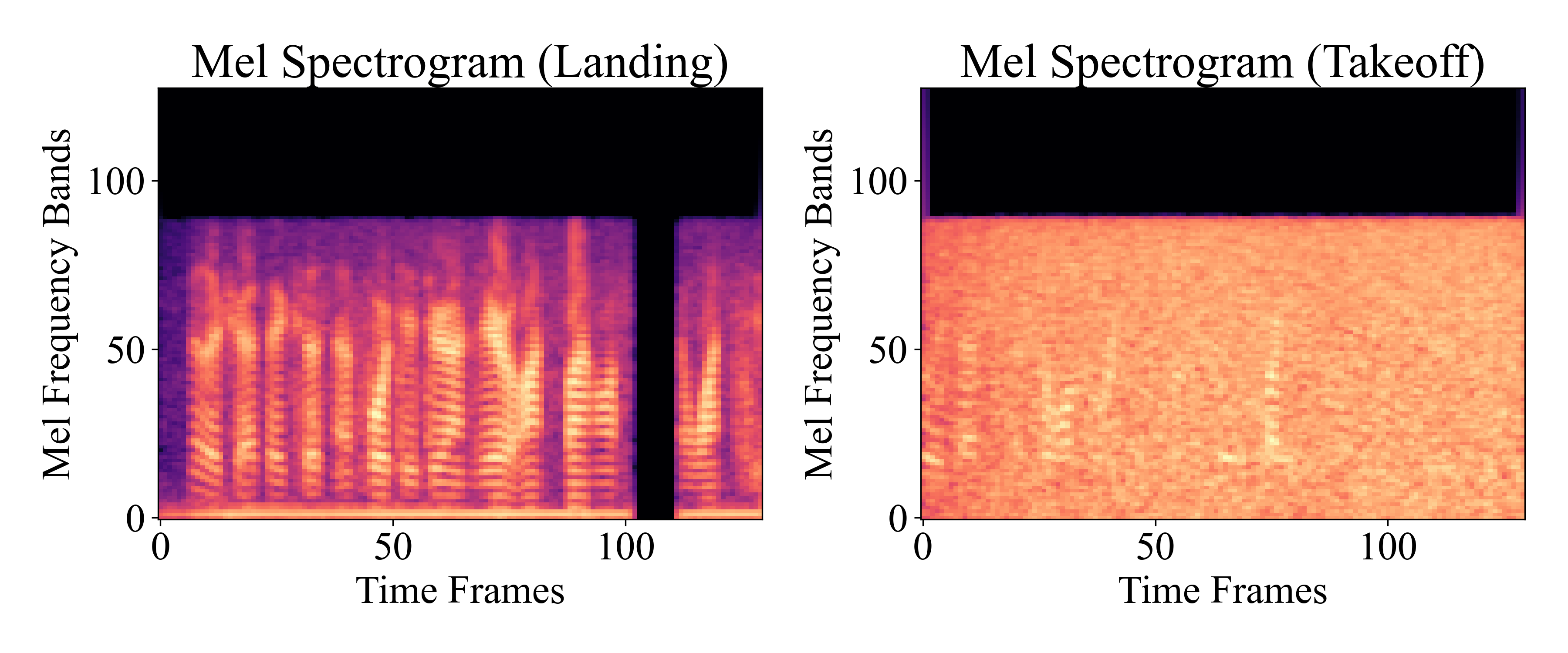}
    \vspace{-1.5em}
    \caption{Examples of Mel-spectrograms for both classes.}
    \label{fig:mel_spectrogram}
    \vspace{-1.0em}
\end{figure}

To maintain uniform input shape compatible with convolutional neural networks, shorter recordings were zero-padded along the time axis and longer ones were truncated. An illustration of Mel-spectrograms for both "Landing" and "Takeoff" classes is shown in Fig.~\ref{fig:mel_spectrogram}.

Following feature extraction, valid spectrograms were collected into a NumPy array with an additional channel dimension, resulting in a dataset of shape \( (N, 128, 130, 1) \), where \( N \) is the number of audio samples. This ensured compatibility with 2D convolutional neural network architectures for subsequent classification tasks.


\subsection{\textbf{Audio Data Augmentation}}
\label{subsec:augmentation}

To improve model robustness and generalization, we applied audio augmentation techniques during training to simulate real-world variations in pilot speech and acoustic conditions. Specifically, we used three methods: time stretching (10\% speed increase without pitch change) to mimic varying speech rates, Gaussian noise injection (noise factor 0.005) to replicate ambient sounds like wind or static, and temporal shifting (up to 10\% of duration) to account for speech timing differences. These augmentations preserved the semantic content of the audio and were applied only during training, with test data left unchanged to ensure fair evaluation.

\begin{algorithm}[!th]
\caption{Audio Classification via Textual and Spectral Feature Pipelines}
\label{alg:audio_classification}
\begin{algorithmic}[1]
\REQUIRE Audio dataset \( \mathcal{D} = \{(x_i, y_i)\}_{i=1}^{N} \), where \( x_i \) is a waveform and \( y_i \in \{\text{Landing}, \text{Takeoff}\} \)
\ENSURE Predicted labels \( \hat{y}_i \) for each \( x_i \)

\FOR{each \( x_i \in \mathcal{D} \)}
    \STATE Preprocess audio:
    \begin{itemize}[leftmargin=*, noitemsep, topsep=0pt]
        \item \( x_i^{\text{mono}} \gets \text{Mono}(x_i) \)
        \item \( P_n(f) \gets \text{PSD}(x_i^{\text{mono}}[0:T_n]) \)
        \item \( X_i(t, f) \gets \text{STFT}(x_i^{\text{mono}}) \)
        \item \( \hat{X}_i(t, f) \gets \max(|X_i(t, f)| - P_n(f), 0) \)
        \item \( \hat{x}_i(t) \gets \text{ISTFT}(\hat{X}_i(t, f), \angle X_i(t, f)) \)
    \end{itemize}
    
    \STATE Extract features:
    \begin{itemize}[leftmargin=*, noitemsep, topsep=0pt]
        \item \textbf{Textual:}
        \begin{itemize}[leftmargin=1.5em, noitemsep]
            \item \( s_i \gets \text{ASR}(\hat{x}_i) \)
            \item \( \mathbf{v}_i^{\text{text}} \gets \text{TFIDF}(s_i) \)
        \end{itemize}
        \item \textbf{Spectral:}
        \begin{itemize}[leftmargin=1.5em, noitemsep]
            \item \( \tilde{x}_i \gets \text{Pad}(\text{Resample}(\hat{x}_i), 3\text{s}) \)
            \item \( M_i \gets \text{MelSpec}(\tilde{x}_i; \text{FFT}=2048, \text{Bands}=128) \)
            \item \( \mathbf{v}_i^{\text{spec}} \gets \text{Normalize}(\log M_i) \in \mathbb{R}^{128 \times 130} \)
        \end{itemize}
    \end{itemize}
\ENDFOR

\STATE Train classifier \( f: \mathbf{v}_i \rightarrow \hat{y}_i \), where:
\[
\mathbf{v}_i =
\begin{cases}
\mathbf{v}_i^{\text{text}} & \text{Textual pipeline} \\
\mathbf{v}_i^{\text{spec}} & \text{Spectral pipeline}
\end{cases}
\]

\STATE Evaluate \( \{\hat{y}_i\}_{i=1}^{N} \) using accuracy, precision, recall, and F1-score
\end{algorithmic}
\end{algorithm}

\subsection{\textbf{Classification Models and Training Procedure}}
\label{subsec:algorithm}

To evaluate the Textual and Spectral pipelines, we implemented a diverse set of models \( \mathcal{M} = \{M_1, M_2, \dots, M_k\} \), including both traditional classifiers—Logistic Regression, Decision Tree, Random Forest, Support Vector Machine, K-Nearest Neighbors, and Gradient Boosting—and deep learning architectures (CNN and LSTM). These models were trained and tested on consistent data splits across both pipelines. CNNs were applied to 2D Mel-spectrograms, while LSTMs processed ASR-transcribed text sequences. Additionally, we incorporated an ensemble model using soft voting, where the final prediction is given by:
\vspace{-0.5em}
\begin{equation}
\hat{y}_{\text{ensemble}} = \arg\max_{j} \sum_{i=1}^{n} p_{i,j}
\label{eq:softvoting}
\end{equation}

\noindent
Here, \( p_{i,j} \) denotes the probability assigned to class \( j \) by model \( M_i \). All hyperparameters were tuned via grid search or manual optimization, enabling consistent benchmarking across architectures.

\section{Experimental Analysis}
\label{sec:exp_analysis}

\begin{table*}[!h]
\centering
\caption{Performance of Traditional and Deep Learning Models Across Textual and Spectral Pipelines}
\label{tab:all_model_performance}
\begin{tabular}{lcccccc|cccccc}
\toprule
\multirow{2}{*}{\textbf{Model}} & \multicolumn{6}{c|}{\textbf{Textual (TF-IDF)}} & \multicolumn{6}{c}{\textbf{Spectral (Mel-Spectrogram)}} \\
\cmidrule(lr){2-7} \cmidrule(lr){8-13}
 & Acc. & Prec. & Rec. & F1 & AUROC & AUPR & Acc. & Prec. & Rec. & F1 & AUROC & AUPR \\
\midrule
Logistic Regression       & 0.82 & 0.81 & 0.80 & 0.80 & 0.85 & 0.84 & 0.85 & 0.84 & 0.83 & 0.83 & 0.88 & 0.87 \\
Support Vector Machine    & 0.83 & 0.82 & 0.82 & 0.82 & 0.86 & 0.85 & 0.87 & 0.86 & 0.85 & 0.86 & 0.90 & 0.89 \\
K-Nearest Neighbors       & 0.78 & 0.77 & 0.76 & 0.76 & 0.82 & 0.81 & 0.80 & 0.79 & 0.78 & 0.78 & 0.83 & 0.82 \\
Random Forest             & 0.84 & 0.83 & 0.83 & 0.83 & 0.87 & 0.86 & 0.89 & 0.88 & 0.87 & 0.88 & 0.91 & 0.90 \\
Gradient Boosting         & 0.85 & 0.84 & 0.84 & 0.84 & 0.88 & 0.87 & 0.90 & 0.89 & 0.89 & 0.89 & 0.93 & 0.92 \\
Ensemble Voting           & \textbf{0.86} & \textbf{0.85} & \textbf{0.85} & \textbf{0.85} & \textbf{0.89} & \textbf{0.88} & 0.88 & 0.89 & 0.88 & 0.88 & 0.90 & 0.91 \\
LSTM (Deep Learning)      & 0.84 & 0.83 & \textbf{0.85} & 0.84 & 0.88 & 0.86 & --   & --   & --   & --   & --   & --   \\
CNN (Deep Learning)       & --   & --   & --   & --   & --   & --   & \textbf{0.93} & \textbf{0.91} & \textbf{0.92} & \textbf{0.91} & \textbf{0.95} & \textbf{0.94} \\
\bottomrule
\end{tabular}
\vspace{-1.0em}
\end{table*}

This section presents a detailed evaluation of the proposed audio classification framework, comparing the Textual and Spectral pipelines using various machine learning and deep learning models. The dataset is split 80\%-20\% for training and testing, with deep models trained using the Adam optimizer (learning rate 0.001) and Binary Crossentropy loss.

The results are organized as follows: Subsections~\ref{subsec:model_performance} detail model-wise performance for each pipeline; Subsection~\ref{subsec:robustness} presents robustness results via augmentation; Subsection~\ref{subsec:feature_extraction} analyzes feature extraction techniques; and Subsection~\ref{subsec:correlation} discusses metric correlations.

\subsection{\textbf{Model Performance Across Pipelines}}
\label{subsec:model_performance}

To evaluate model performance across different learning paradigms, we benchmarked six traditional classifiers and two deep learning models on both the textual (TF-IDF) and spectral (Mel-spectrogram) pipelines. Table~\ref{tab:all_model_performance} presents the results for all models across six metrics. Overall, models using spectral features consistently outperform their textual counterparts. Among traditional classifiers, Gradient Boosting and Random Forest achieved strong and balanced results across all metrics, especially in the spectral pipeline. The CNN model outperformed all others with the highest AUROC (0.95) and AUPR (0.94), highlighting the effectiveness of deep learning with time-frequency features for audio classification.

\subsection{\textbf{Feature Representations and Comparison}}
\label{subsec:feature_extraction}
\vspace{-0.25em}
To evaluate the effect of different feature representations, we conducted an ablation study comparing TF-IDF and BERT embeddings for textual inputs, and Mel versus Log-Mel spectrograms for spectral inputs. As shown in Table~\ref{tab:feature_comparison}, TF-IDF outperformed BERT across most traditional classifiers, while BERT showed a slight advantage with the LSTM model. In the spectral pipeline, Log-Mel features yielded modest improvements for Gradient Boosting and Ensemble models, though standard Mel-spectrograms remained more effective for CNNs. These results suggest that the choice of feature representation should be tailored to both the model architecture and the nature of the input data.

\begin{table}[!h]
\centering
\caption{Accuracy with Different Feature Representations}
\label{tab:feature_comparison}
\scriptsize 
\begin{tabular}{lcc|cc}
\toprule
\textbf{Model} & \textbf{TF-IDF} & \textbf{BERT} & \textbf{Mel} & \textbf{Log-Mel} \\
              & \textbf{(Textual)} & \textbf{(Textual)} & \textbf{(Spectral)} & \textbf{(Spectral)} \\
\midrule
LR                & 0.82 & 0.78 & 0.85 & 0.84 \\
SVM               & 0.83 & 0.80 & 0.87 & 0.84 \\
KNN               & 0.78 & 0.77 & 0.80 & 0.81 \\
RF                & 0.84 & \textbf{0.86} & 0.89 & 0.88 \\
GB                & 0.85 & 0.82 & 0.90 & \textbf{0.92} \\
EV                & \textbf{0.86} & 0.79 & 0.88 & 0.90 \\
LSTM (Textual)    & 0.84 & 0.85 & --   & --   \\
CNN (Spectral)    & --   & --   & \textbf{0.93} & 0.89 \\
\bottomrule
\end{tabular}
\end{table}

\subsection{\textbf{Robustness Analysis via Data Augmentation}}
\label{subsec:robustness}

To evaluate model robustness and generalization, we applied audio data augmentation during training for all models, both traditional machine learning and deep learning, across the Textual and Spectral pipelines. These augmentations simulate realistic variations in pilot speech and environmental noise, enabling models to better generalize to unseen data.

The augmentation techniques included time stretching (factor = 1.1), additive Gaussian noise (noise factor = 0.005), and temporal shifting (up to 10\% of audio duration). All augmentations were applied exclusively during training. Test data remained unmodified to ensure fair evaluation.

\begin{figure}[!th]
    \centering
    \includegraphics[width=\columnwidth]{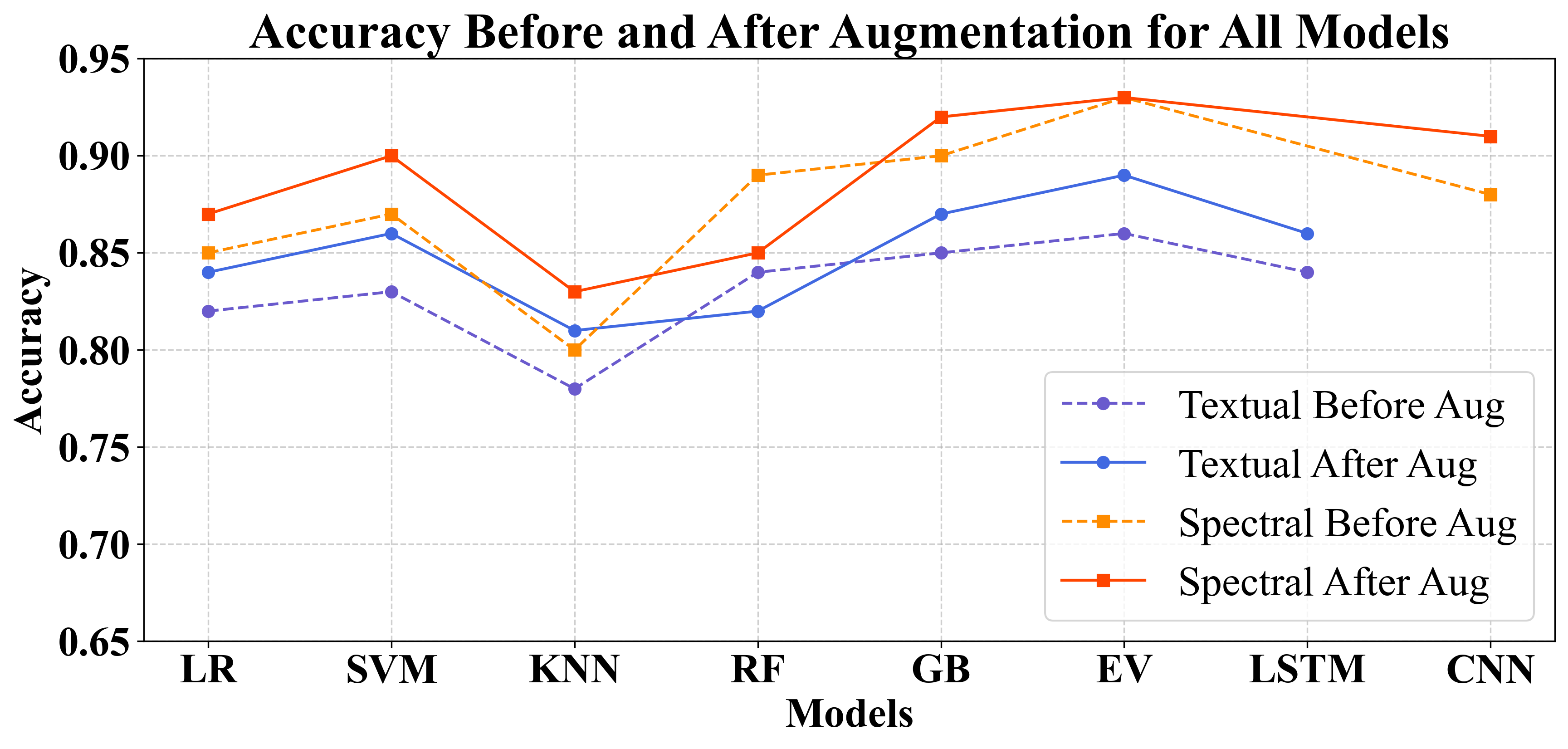}
    \caption{Accuracy comparison of models before and after audio data augmentation.}
    \vspace{-0.5em}
    \label{fig:robustness_line}
    \vspace{-0.5em}
\end{figure}

Figure~\ref{fig:robustness_line} summarizes the performance impact of augmentation, showing results from representative models in each pipeline. Notably, all models benefited from augmentation, with improvements observed across accuracy, F1-Score, AUROC, and AUPR.

These results demonstrate that augmentation significantly improves model performance across both feature modalities and model types. Spectral models, particularly CNNs, benefited the most, but consistent gains were also observed in textual models, underscoring the value of training with varied and noisy inputs.


\subsection{\textbf{Evaluation Metric Correlation Analysis}}
\label{subsec:correlation}
\vspace{-0.25em}

\begin{figure*}[!h]
    \centering
    \begin{subfigure}[t!]{0.48\textwidth}
        \centering
        \includegraphics[width=0.9\linewidth]{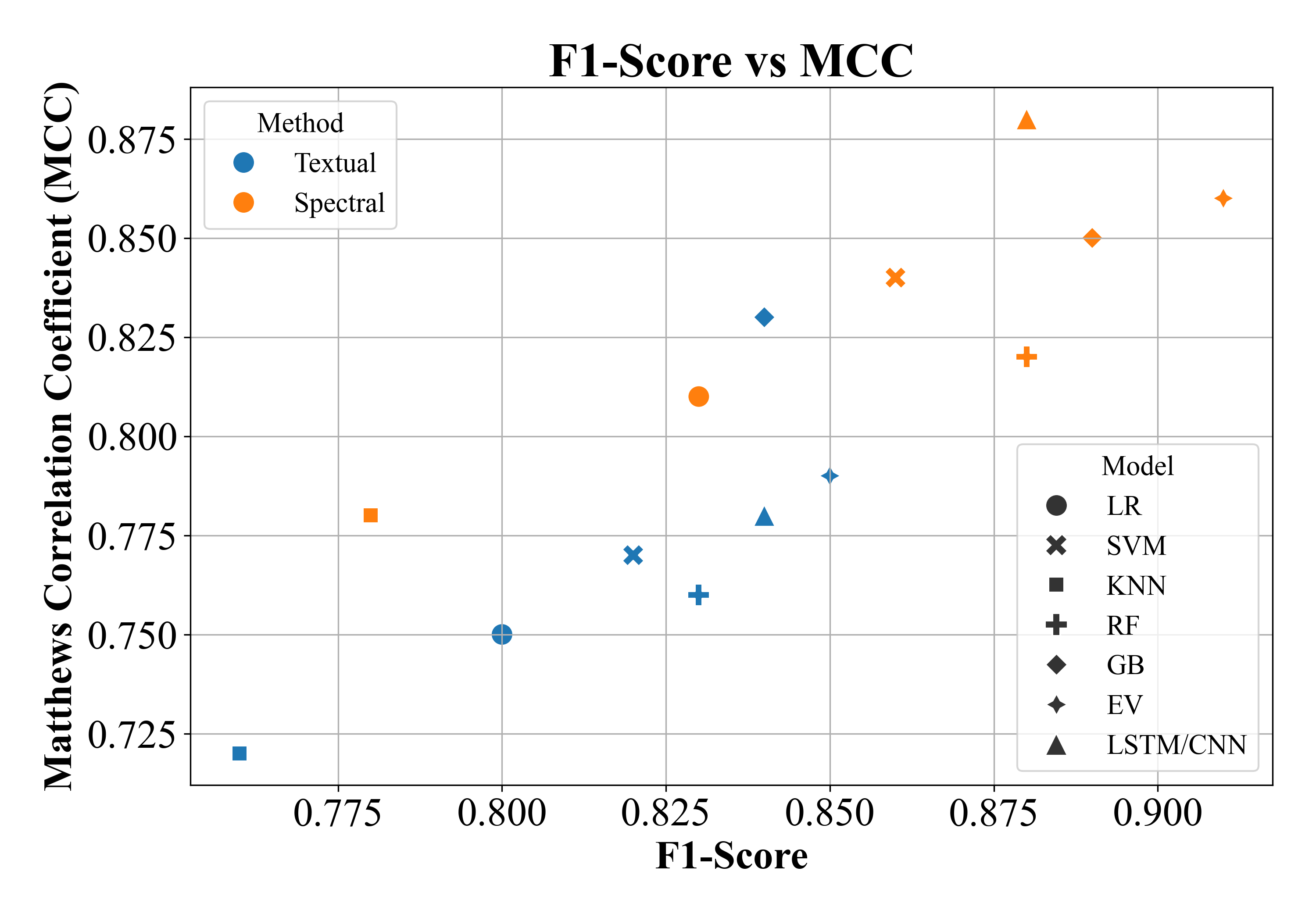}
        \caption{F1-Score vs. MCC}
        \label{fig:f1_mcc}
    \end{subfigure}
    \hfill
    \begin{subfigure}[t!]{0.48\textwidth}
        \centering
        \includegraphics[width=0.9\linewidth]{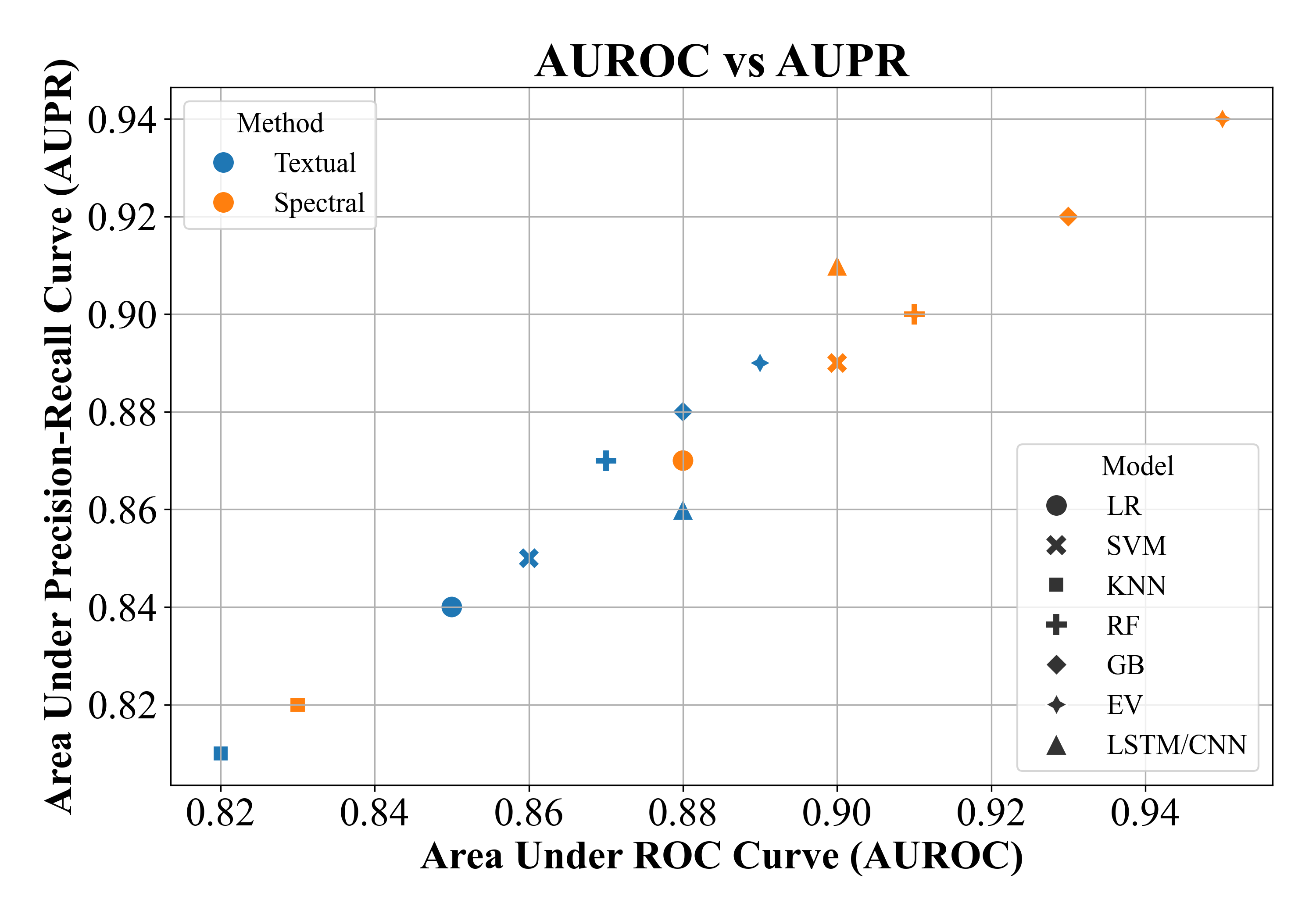}
        \caption{AUROC vs. AUPR}
        \label{fig:auroc_aupr}
    \end{subfigure}
    \caption{Metric correlation plots for all models across both pipelines.}
    \label{fig:metric_correlation}
    \vspace{-0.5em}
\end{figure*}

To better understand model performance, we examine the correlation between key evaluation metrics across all configurations, as shown in Figure~\ref{fig:metric_correlation}. The F1-Score vs. MCC plot reveals a strong positive relationship, with Spectral models, especially CNN which is clustered in the upper-right, indicating consistent and balanced predictions. Similarly, the AUROC vs. AUPR plot shows that models with high AUROC also achieve strong precision-recall performance. These trends highlight the robustness and generalizability of models trained on Mel-spectrogram features.

\section{Conclusion}
\label{sec:conclusion}
\vspace{-0.3em}
In this study, we proposed a dual-pipeline approach for classifying air traffic communication audio into “Landing” and “Takeoff” categories, leveraging both textual and spectral representations. The Textual pipeline utilized automatic speech recognition (ASR) followed by TF-IDF vectorization to capture semantic and operational content, while the Spectral pipeline extracted Mel-spectrograms to retain key acoustic features. A comprehensive set of traditional machine learning and deep learning models was evaluated across both pipelines, including an ensemble classifier that aggregated predictions via soft voting. To improve model robustness, we applied audio augmentations such as time stretching, noise injection, and temporal shifting. These techniques notably enhanced classification accuracy, especially for deep learning models, and increased resilience to speech and noise variations.

Our findings suggest that combining multiple feature representations with appropriate modeling and augmentation strategies can lead to effective and scalable solutions for real-world aviation communication tasks. The proposed framework requires no additional hardware and is well-suited for deployment at both towered and non-towered airports, making it a practical and cost-effective tool for future air traffic monitoring systems.





\vspace{-1.0em}
\bibliographystyle{IEEEtran}
\bibliography{reference}

\end{document}